\documentclass[onefignum,onetabnum,11pt]{siamart190516}

\usepackage{amssymb}
\usepackage{tikz}
\usepackage{wrapfig}

\newcommand{\X}{{\bf x}}
\newcommand{\Y}{{\bf y}}

\newcommand{\R}{\mathbb{R}}
\newcommand{\Z}{{\bf z}}

\newcommand{\fma}{{\small \tt fma}}

\headers{Complexity of the Chain Rule}{}

\title{On the Computational Complexity of the Chain Rule of Differential Calculus}
\author{Uwe Naumann\thanks{Informatik 12: Software and Tools for Computational Engineering, RWTH Aachen University, D-52056 Aachen, Germany (\email{naumann@stce.rwth-aachen.de}, \url{http://www.stce.rwth-aachen.de})}}

\begin{document}

\maketitle

\begin{abstract}
Many modern numerical methods in computational science and engineering rely 
on derivatives of mathematical models for the phenomena under investigation.
The computation of these derivatives often represents the bottleneck
in terms of overall runtime performance. First and higher derivative 
tensors need to be evaluated efficiently.

The chain rule of differentiation is the fundamental prerequisite for
computing accurate derivatives of composite functions which perform a potentially very large number of elemental function evaluations. Data flow dependences amongst the elemental functions give rise to a combinatorial  optimization
problem.
We formulate {\sc Chain Rule Differentiation} and we prove it to 
be NP-complete. Pointers to research on its approximate solution are given. 
\end{abstract}



\section{Introduction}

The chain rule is a classic of differential calculus. Hence, it is all the more
surprising that first successful steps towards a rigorous computational 
complexity analysis were taken only in 2008 \cite{Naumann2008OJa}. A proof
of NP-completeness of {\sc [Optimal] Jacobian Accumulation} was presented 
which is generalized in this paper for derivatives of arbitrary order.

In its simplest form, the chain rule of differentiation reads as
\begin{equation} \label{eqn:cr}
	F'=\prod_{i=1}^{q} F'_i \equiv F'_q \cdot F'_{q-1} \cdot \ldots \cdot F'_1 \; .
\end{equation}
We use $=$ to denote equality and $\equiv$ is the sense of ``is defined as.''
Differentiability of the {\em elemental functions} 
\begin{equation} \label{eqn:Fi}
F_i : \R^{n_{i-1}} \rightarrow \R^{n_i} \; : \quad \Z_{i-1} \mapsto \Z_i=F_i(\Z_{i-1})
\end{equation}
for $n_0=n$ and $n_q=m$ implies differentiability of the {\em composite function}
\begin{equation} \label{eqn:F}
F : \R^n \rightarrow \R^m \; : \quad \X \mapsto \Y=F(\X)=F_q(F_{q-1}(\ldots F_1(\X) \ldots ))
\end{equation}
where $\Z_0=\X$ and $\Y=\Z_q.$ 
Vectors are printed in bold type.
The matrix chain product of given {\em elemental 
Jacobians}
\begin{equation} \label{eqn:Fi'}
F'_i=F'_i(\Z_{i-1}) \equiv \frac{\partial F_i}{\partial \Z_{i-1}}(\Z_{i-1}) \in \R^{n_i \times n_{i-1}}
\end{equation}
yields  
\begin{equation} \label{eqn:F'}
F'=F'(\X) \equiv \frac{\partial F}{\partial \X}(\X) \in \R^{m \times n} \; ,
\end{equation}
where $\frac{\partial C}{\partial D}$ denotes the (partial) derivative of 
the counter $C$ with respect to the denominator $D.$

Associativity of matrix multiplication gives rise to the 
{\sc Jacobian Chain Bracketing} problem which can be solved by dynamic
programming \cite{Bellman1957DP,Godbole1973} if the scalar entries of all 
$F'_i$ are assumed to be 
algebraically independent. Potential equality of entries makes the 
general {\sc Jacobian Accumulation} problem NP-complete as shown in
\cite{Naumann2008OJa}, where the same problem is referred to as 
{\sc Optimal Jacobian Accumulation}. The proof uses 
reduction \cite{Karp1972RAC} from 
{\sc Ensemble Computation} \cite{Garey1979CaI}. An extension of the idea is used in this paper 
for the generalization to derivatives of arbitrary order. 

Numerical simulations in computational science and engineering often result
in composite functions of highly complex structure.
Algorithmic differentiation (AD) \cite{Griewank2008EDP,Naumann2012TAo,Rall1981ADT,Wengert1964ASA} 
(also known as automatic differentiation or differentiable programming) 
is applicable to differentiable multivariate vector functions 
$F : \R^n \rightarrow \R^m$ implementing
\begin{equation} \label{eqn:sac}
\begin{split}
\Z_0&=\X \\
\Z_j&=F_j\left ((\Z_i)_{i \prec j} \right) \quad \text{for}~j=1,\ldots,q \\
\Y&=\Z_q \; ,
\end{split}
\end{equation}
where, adopting notation from \cite{Griewank2008EDP}, $i \prec j$ if and only 
if $\Z_i$ is an argument of $F_j.$ A directed acyclic graph (dag) $G=(V,E)$ with
integer vertices $V=\{0,\ldots,q\}$ and edges $E=\{(i,j):i \prec j\}$ is induced.
Elemental Jacobians $F'_{j,i}$ are associated with all edges $(i,j) \in E.$
The (first-order) chain rule becomes 
\begin{equation} \label{eqn:cr1}
F'=\sum_{(0,\ldots,q)} \; \prod_{(i,j) \in (0,\ldots,q)} F'_{j,i} 
\end{equation} 
\cite{Baur1983TCo}. 
Equation~(\ref{eqn:F}) represents a special case. 
Summation is over all paths $(0,\ldots,q)$ in $G$ 
connecting the input 
$\Z_0=\X$ with the output $\Y=\Z_q.$ Products of elemental Jacobians along 
all paths are evaluated as in Equation~(\ref{eqn:cr}).
An example is shown in Figure~\ref{fig:1}.
\begin{figure}
	\begin{minipage}[c]{.75\linewidth}
		$$
		\Y=F(\X)=F_3(F_2(F_1(\X),\X),F_1(\X)) \quad \Rightarrow
		$$
		$\;$ \vspace{-.5cm} \\
		$$
		\Downarrow
		$$
	\end{minipage}
	\begin{minipage}[c]{.15\linewidth}
		\centering
	\begin{tikzpicture}[scale=.6, transform shape]
  \begin{pgfscope}
    \tikzstyle{every node}=[draw,circle,minimum size=1cm]
          \node (0) at (1,0) {$0$};
          \node (1) at (0,1) {$1$};
          \node (2) at (2,2) {$2$};
          \node (3) at (1,3) {$3$};
  \end{pgfscope}
 \begin{scope}[-latex]
 \draw (0) -- (1);
 \draw (0) -- (2);
 \draw (1) -- (2);
 \draw (2) -- (3);
 \draw (1) -- (3);
  \end{scope}
\end{tikzpicture} 
	\end{minipage}

	\begin{minipage}[c]{\linewidth}
		$$
			F'=F'_{3,1}(\Z_1,\Z_2) \cdot F'_{1,0}(\Z_0)+
			F'_{3,2}(\Z_1,\Z_2) \cdot F'_{2,1}(\Z_0,\Z_1) \cdot F'_{1,0}(\Z_0)+
			F'_{3,2}(\Z_1,\Z_2) \cdot F'_{2,0}(\Z_0,\Z_1)
			$$
	\end{minipage}
	\caption{Illustration of Equation~(\ref{eqn:cr1})} \label{fig:1}
\end{figure}

For given elemental Jacobians
$$
F'_{j,i} = F'_{j,i}\left ((\Z_{k})_{k \prec i} \right) \equiv
\frac{\partial F_j}{\partial \Z_i}\left ((\Z_{k})_{k \prec i} \right) 
$$
the evaluation of Equation~(\ref{eqn:cr1}) breaks down into a sequence of
fused-multiply-add (\fma) operations, where
each scalar multiplication is optionally followed by a scalar
addition.
Higher-order chain rules follow naturally. For example, the second-order 
chain rule for composite functions as in Equation~(\ref{eqn:F}) becomes
\begin{equation} \label{cr2}
\left [F'' \right ]_{\delta,\alpha_1,\alpha_2}=\sum_{j=1}^q \left ( \left [\prod_{i=j+1}^q F'_i \right ]_{\delta,\gamma} \left [F''_j \right ]_{\gamma,\beta_1,\beta_2} \left [\prod_{k=1}^{j-1} F'_k \right ]_{\beta_1,\alpha_1} \left [\prod_{k=1}^{j-1} F'_k \right ]_{\beta_2,\alpha_2} \right )
\end{equation} 
It
describes the computation of the Hessian tensor 
$F'' = \left [F'' \right ]_{\delta,\alpha_1,\alpha_2} \in \R^{m \times n \times n}$ for given elemental Jacobians and Hessians.
Index notation (summation over the shared index) is used.
The corresponding tensors are enclosed in square brackets.
An example can be found in Figure~\ref{fig:2}.

We are interested in minimizing the number of \fma\ required to evaluate the
$p$-th-order chain rule for $p=1,2,\ldots$.
The indexing in third- and higher-order chain rules becomes rather involved.
The corresponding formulas are omitted as 
they are not required for the following argument. 

\section{Complexity Analysis}

The complexity analysis is conducted for the following decision problem.
\begin{definition}[\sc Chain Rule Differentiation] \begin{itemize}
\item[] 
	{\em INSTANCE:} A composite function as in Equation~(\ref{eqn:sac}) with given elemental derivatives up to order $p$ and a positive integer $K$. 
\item[] 
	{\em QUESTION:} Can the $p$-th derivative of $F$ be computed with at most $K$ \fma\ operations?
\end{itemize}
\end{definition}
Gradual decrease of feasible $K$ yields solutions to the corresponding optimization
problem.

{\sc Chain Rule Differentiation} turns out to be NP-complete.
The proof uses reduction from the following combinatorial problem.
\begin{definition}[\sc Ensemble Computation] \begin{itemize}
\item[] 
	{\em INSTANCE:} A collection
$C = \{C_\nu \subseteq A : \nu=1,\ldots,|C|\}$ of subsets
$C_\nu = \{c_i^\nu:i=1,\ldots,|C_\nu|\}$
of a finite set $A$ and a positive integer
$K.$
\item[] 
	{\em QUESTION:} Is there a sequence
$u_i=s_i \cup t_i$ for $i=1,\ldots,k$ of $k \leq K$ union
operations, where each $s_i$ and $t_i$ is either $\{a\}$ for some $a \in A$
or $u_j$
for some
$j < i,$ such that $s_i$ and $t_i$ are disjoint for $i=1,\ldots,k$ and
such that for every subset $C_\nu \in C,$ $\nu=1,\ldots,|C|,$
there is some $u_i,$ $1 \leq i \leq k,$ that is identical to $C_\nu?$
\end{itemize}
\end{definition}
Instances of {\sc Ensemble Computation} are given as triplets $(A,C,K).$
For example, for $A=\{a_1,a_2,a_3,a_4\},$ $C=\left \{\{a_1,a_2\},\{a_2,a_3,a_4\},\{a_1,a_3,a_4\}\right \}$ and $K=4$ the answer 
is positive as
$C_1=u_1=\{a_1\} \cup \{a_2\};$ $u_2=\{a_3\} \cup \{a_4\};$ $C_2=u_3=\{a_2\} \cup u_2;$ $C_3=u_4=\{a_1\} \cup u_2.$
$K=3$ yields a negative answer identifying $K=4$ as the solution of
the corresponding optimization problem.
\begin{lemma}
{\sc Ensemble Computation} is NP-complete. 
\end{lemma}
\begin{proof}
See \cite{Garey1979CaI}.
\end{proof}
The proof of the following theorem establishes NP-completeness of 
{\sc Chain Rule Differentiation} for derivatives of arbitrary order.
\begin{theorem} \label{the}
{\sc Chain Rule Differentiation} is NP-complete.
\end{theorem}
\begin{figure}
	\begin{minipage}[c]{\linewidth}
		\footnotesize
		\begin{align}
			F^{[1]}&=F^{[1]}_3 \cdot F^{[1]}_2 \cdot F^{[1]}_1 \label{1}\\
			[F^{[2]}]_{\delta,\alpha_1,\alpha_2} &=\underset{=0}{\underbrace{[F^{[2])}_3]_{\delta,\gamma_1,\gamma_2}}} \cdot [F^{[1]}_2 \cdot F^{[1]}_1]_{\gamma_1,\alpha_1} \cdot [F^{[1]}_2 \cdot F^{[1]}_1]_{\gamma_2,\alpha_2} \label{2:1} \\
			&\; + [F^{[1]}_3]_{\delta,\gamma} \cdot \underset{=0}{\underbrace{[F^{[2]}_2]_{\gamma,\beta_1,\beta_2}}} \cdot [F^{[1]}_1]_{\beta_1,\alpha_1} \cdot [F^{[1]}_1]_{\beta_2,\alpha_2} \label{2:2} \\
			&\; + [F^{[1]}_3 \cdot F^{[1]}_2]_{\delta,\beta} \cdot [F^{[2]}_1]_{\beta,\alpha_1,\alpha_2} \label{2:3} \\
			[F^{[3]}]_{\delta,\alpha_1,\alpha_2,\alpha_3} &=\underset{=0}{\underbrace{[F^{[3]}_3]_{\delta,\gamma_1,\gamma_2,\gamma_3}}} \cdot [F^{[1]}_2 \cdot F^{[1]}_1]_{\gamma_1,\alpha_1} \cdot [F^{[1]}_2 \cdot F^{[1]}_1]_{\gamma_2,\alpha_2} \cdot [F^{[1]}_2 \cdot F^{[1]}_1]_{\gamma_3,\alpha_3} \label{3:1} \\
			&\;+\underset{=0}{\underbrace{[F^{[2]}_3]_{\delta,\gamma_1,\gamma_2}}} \cdot \ldots \quad \text{remaining terms due to (\ref{2:1})}\\
			&\; + \underset{=0}{\underbrace{[F^{[2]}_3]_{\delta,\gamma_1,\gamma_2}}} \cdot \underset{=0}{\underbrace{[F^{[2]}_2]_{\gamma_1,\beta_1,\beta_2}}} \cdot [F^{[1]}_1]_{\beta_1,\alpha_1} \cdot [F^{[1]}_1]_{\beta_2,\alpha_2} \cdot [F^{[1]}_2]_{\gamma_2,\alpha_3}\\
			&\; + [F^{[1]}_3]_{\delta,\gamma} \cdot \underset{=0}{\underbrace{[F^{[3]}_2]_{\gamma,\beta_1,\beta_2,\beta_3}}} \cdot [F^{[1]}_1]_{\beta_1,\alpha_1} \cdot [F^{[1]}_1]_{\beta_2,\alpha_2} \cdot [F^{[1]}_1]_{\beta_3,\alpha_3}\\
			&\; + [F^{[1]}_3]_{\delta,\gamma} \cdot \underset{=0}{\underbrace{[F^{[2]}_2]_{\gamma,\beta_1,\beta_2}}} \cdot \ldots \quad \text{remaining terms due to (\ref{2:2})} \\
			&\; + \underset{=0}{\underbrace{[F^{[2]}_3]_{\delta,\gamma_1,\gamma_2}}} \cdot [F^{[1]}_2]_{\gamma_1,\beta} \cdot [F^{[2]}_1]_{\beta,\alpha_1,\alpha_2} \cdot [F^{[1]}_2 \cdot F^{[1]}_1]_{\gamma_2,\alpha_3} \\
			&\; + [F^{[1]}_3]_{\delta,\gamma} \cdot \underset{=0}{\underbrace{[F^{[2]}_2]_{\gamma,\beta_1,\beta_2}}} \cdot [F^{[2]}_1]_{\beta_1,\alpha_1,\alpha_2} \cdot [F^{[1]}_2]_{\beta_2,\alpha_3} \\
			&\; + [F^{[1]}_3 \cdot F^{[1]}_2]_{\delta,\beta} \cdot [F^{[3]}_1]_{\beta,\alpha_1,\alpha_2,\alpha_3} \label{3:2}
		\end{align}
	\end{minipage}
	\caption{Illustration of induction in proof of Theorem~\ref{the} for $q=3$ and $p=1,2,3$}
	\label{fig:2}
\end{figure}
\begin{proof}
Consider an arbitrary instance $(A,C,K)$ of {\sc Ensemble Computation} and a bijection
$A \leftrightarrow \tilde{A},$ where $\tilde{A}$ consists of $|A|$ mutually
	distinct primes\footnote{The proof in \cite{Naumann2008OJa} does not mention primes explicitly. However, their use in connection with the uniqueness
	property due to the 
fundamental theorem of arithmetic \cite{Gauss1801DA} turns out to be crucial
	for the correctness of the overall argument.} $\in \{2,3,5,\ldots\}.$
A corresponding bijection
$C \leftrightarrow \tilde{C}$ is implied.
Create an extended version $(\tilde{A} \cup \tilde{B},\tilde{C},K+|\tilde{B}|)$
of {\sc Ensemble Computation}
by adding unique entries from a sufficiently large set $\tilde{B}$
of primes not in $\tilde{A}$ to the $\tilde{C}_j$ such that they end up having
the same cardinality $q$. Note that a solution for this extended
instance of {\sc Ensemble Computation} implies a solution of the original instance as each entry of $\tilde{B}$ appears exactly once.

Fix the order of the elements of the
$\tilde{C}_j$ arbitrarily yielding
$\tilde{C}_j=(\tilde{c}^j_i)_{i=1}^q$ for $j=1,\ldots,|\tilde{C}|.$
Let
$$
F : \R \rightarrow \R^{|\tilde{C}|} : \quad \Y=\Z_q=F(x)
$$
	with $F(x)=F_q( F_{q-1} ( \ldots F_1(x) \ldots ))$
defined as
\begin{align*}
F_1 &: \R \rightarrow \R^{|\tilde{C}|} : \quad \Z_1=F_1(x) : \;
z^1_j=\frac{\tilde{c}^j_1}{p!} \cdot x^p 
	\intertext{and}
F_i &: \R^{|\tilde{C}|} \rightarrow \R^{|\tilde{C}|} : \quad \Z_i=F_i(\Z_{i-1}) :   \;
z^i_j=\tilde{c}^j_i \cdot z^{i-1}_j \; ,
\end{align*}
	where $\Z_i=(z^i_j).$ 
The $p$-th derivative of $F_1$ becomes equal to
$$
	F^{[p]}_1=\left (\tilde{c}^j_1 \right) \in \R^{|\tilde{C}|}=\R^{{|\tilde{C}|} \times 1 \times \ldots (p~\text{times}) \ldots \times 1} \; .
$$
The remaining 
	Jacobians ($i=2,\ldots,q$) $$F^{[1]}_i=F'_i=(d^i_{j,k}) \in \R^{{|\tilde{C}| \times |\tilde{C}|}} \; ,$$
	turn out to be diagonal matrices with
$$
d^i_{j,k}=
\begin{cases}
\tilde{c}^j_i & \text{if}~j=k \\
0 & \text{otherwise} \\
\end{cases}
$$
for $j=1,\ldots,|\tilde{C}|.$ 
From $F^{[p]}_i=0$ for $p>1$ it follows that 
the chain rule of order $p$ simplifies to
	\begin{equation} \label{cr3}
		F^{[p]}=\prod_{i=2}^{q} F_{i}^{[1]} \cdot F_{1}^{[p]} 
	\end{equation} 
which follows by induction over $p$: Obviously, the claim holds for $p=1.$
Refer to Figure~\ref{fig:2}, Equation~(\ref{1}) for illustration for $q=3.$ 

For $p=2$ we get Equation~(\ref{cr2}),
where all terms with $j>1$ vanish identically as they
		contain $F^{[2]}_j=F''_{j}=0$ as a factor.
The remaining term (for $j=1$) yields Equation~(\ref{cr3}).
See Figure~\ref{fig:2}, Equations~(\ref{2:1})--(\ref{2:3}) for illustration.

Suppose that the claim holds for $p-1,$ that is,
$$
		F^{[p-1]}=\prod_{i=2}^{q} F_{i}^{[1]} \cdot F_{1}^{[p-1]} \; .
$$
Application of the (first-order) chain rule yields terms 
		containing $F_{i}^{[2]}=0$ due to differentiation of the $F_{i}^{[1]}$ 
for $i=2,\ldots,q.$ They all vanish identically under the given reduction. Only
the last term due to differentiation of $F_{1}^{[p-1]}$ remains.
		It is equal to the right-hand side of Equation~(\ref{cr3}).
Further illustration is provided in
Figure~\ref{fig:2} for $p=3$ yielding Equations~(\ref{3:1})--(\ref{3:2}).

According to the fundamental theorem of arithmetic \cite{Gauss1801DA}
the elements of $\tilde{C}$
correspond to unique (up to commutativity of scalar multiplication)
factorizations of the $|\tilde{C}|$ nonzero entries of $F^{[p]} \in \R^{|\tilde{C}|}.$
This uniqueness property extends to arbitrary subsets
of the $\tilde{C}_j$ considered during the exploration of the search space
of {\sc Chain Rule Differentiation}.
A solution implies a solution of the associated extended
instance of {\sc Ensemble Computation} and, hence, of the original instance.

A proposed solution for {\sc Chain Rule Differentiation} is easily
validated by counting the at
most $|\tilde{C}|\cdot q$ scalar multiplications performed.
\end{proof}

For illustration consider
the extended version of the example presented for
{\sc Ensemble Computation}:
\begin{align*}
&A=\{a_1,a_2,a_3,a_4\} \Rightarrow \tilde{A}=\{2,3,5,7\} \\
&\tilde{B}=\{11\} \\
&C=\{\{a_1,a_2\},\{a_2,a_3,a_4\},\{a_1,a_3,a_4\}\} 
\Rightarrow \tilde{C}=\left \{\{2, 3, 11\},\{3, 5, 7\}, \{2, 5, 7\}\right \} \\
&K+|\tilde{B}|=K+1=5\; .
\end{align*}
The three nonzero entries of
$$F^{[p]}=F^{[1]}_3 \cdot F^{[1]}_2 \cdot F^{[p]}_1 =
\begin{pmatrix}
11 & & \\
& 7 & \\
&  & 7 \\
\end{pmatrix} 
\cdot
\begin{pmatrix}
3& & \\
& 5& \\
& & 5\\
\end{pmatrix} 
\cdot
\begin{pmatrix}
2 \\
3 \\
2\\
\end{pmatrix} 
$$ 
\begin{wrapfigure}[4]{r}{3cm}
	\centering
	\begin{tikzpicture}[scale=.6, transform shape]
  \begin{pgfscope}
    \tikzstyle{every node}=[draw,circle,minimum size=1cm]
          \node (0) at (1.5,0) {$x$};
          \node (11) at (0,2) {$z^1_1$};
          \node (12) at (1.5,2) {$z^1_2$};
	  \node (13) at (3,2) {$z^1_3$};
          \node (21) at (0,4) {$z^2_1$};
          \node (22) at (1.5,4) {$z^2_2$};
          \node (23) at (3,4) {$z^2_3$};
          \node (31) at (0,6) {$z^3_1$};
          \node (32) at (1.5,6) {$z^3_2$};
          \node (33) at (3,6) {$z^3_3$};
  \end{pgfscope}
 \begin{scope}[-latex]
	 \draw (0) -- (11) node [midway,left] {$2$};
 \draw (0) -- (12) node [midway,left] {$3$};
	 \draw (0) -- (13) node [midway,left] {$2$};
	 \draw (11) -- (21) node [midway,left] {$3$};
	 \draw (12) -- (22) node [midway,left] {$5$};
	 \draw (13) -- (23) node [midway,left] {$5$};
	 \draw (21) -- (31) node [midway,left] {$11$};
	 \draw (22) -- (32) node [midway,left] {$7$};
	 \draw (23) -- (33) node [midway,left] {$7$};
  \end{scope}
\end{tikzpicture} 
	\caption{Example dag} \label{dag2}
\end{wrapfigure}
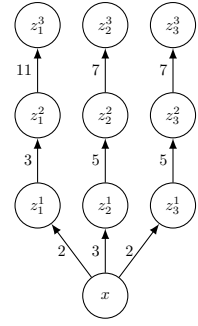
are computed as
\begin{align*}
	\frac{d z^2_1}{d x} &=
\frac{d z^2_1}{d z^1_1} \cdot \frac{d^p z^1_1}{d x^p}=3 \cdot 2 =6 \\ 
	\frac{d z^3_2}{d z^1_2}=\frac{d z^3_3}{d z^1_3}&=
	\frac{d z^3_3}{d z^2_3} \cdot \frac{d z^2_3}{d z^1_3}=
	7 \cdot 5 =35 \\
	F_1^{[p]}&=\tilde{b}_1 \cdot \left (\frac{d z^2_1}{d z^1_1} \cdot \frac{d^p z^1_1}{d x^p} \right )=11 \cdot 6 =66 \\
	F_2^{[p]}&=\frac{d z^3_2}{d z^1_2} \cdot \frac{d^p z^1_2}{d x^p} =35 \cdot 3=105 \\
	F_3^{[p]}&=\frac{d z^3_3}{d z^1_3} \cdot \frac{d^p z^1_3}{d x^p}=35 \cdot 2=70 \; .
\end{align*}
at the expense of five \fma\ (no additions involved)
yielding a positive answer to this instance of the decision version of
{\sc Chain Rule Differentiation}. A corresponding answer to the decision 
version of {\sc Ensemble Computation} is implied. Figure~\ref{dag2}
depicts the corresponding dag.

\section{Tangents and Adjoints}

As an immediate consequence of Theorem~\ref{the} the \fma-optimal evaluation
of tangents and adjoints of arbitrary order turns out to be computationally 
intractable. Tangents and adjoints result from algorithmic
differentiation applied to given implementations of sufficiently often
differentiable multivariate vector functions $\Y=F(\X)$ as in 
Equation~(\ref{eqn:sac}).

The $p$-th-order tangent of $F$ is defined as
$$
\left [\dot{\Y}_p \right ]_k=\left [F^{[p]}(\X)\right ]_{k,j_1,\ldots,j_p} \cdot \prod_{i=1}^{p} \left [ \dot{\X}_i \right ]_{j_i} 
$$
for given input $\X \in \R^n$ and input tangents $\dot{\X}_i \in \R^n.$
It enables the computation of $F^{[p]}$ with a relative (with respect to
the cost of evaluating $F$) computational cost of $\mathcal{O}(n^p)$ 
by letting the
input tangents range independently over the Cartesian basis vectors in $\R^n.$
Exploitation of sparsity is likely to reduce the computational effort 
\cite{Gebremedhin2005WCI}.

The special case in Equation~(\ref{cr3}) resulting from the reduction in the 
proof of Theorem~\ref{the} yields
$$
\dot{\Y}_p=\prod_{i=2}^{q} F_{i}^{[1]} \cdot F_{1}^{[p]} \cdot \prod_{j=1}^{p} \dot{x}_j \; ,
$$
where $\dot{x}_j \in \R$ and $\dot{\Y}_p \in \R^{|\tilde{C}|}.$ A solution for 
$\dot{x}_j=1,$ $j=1,\ldots,p,$ implies a solution of 
{\sc Chain Rule Differentiation}. The corresponding
{\sc Tangent Differentiation} problem is hence at least as hard as 
{\sc Chain Rule Differentiation}. 

Adjoints of order $p$ of $F$ are defined as 
$$
\left [\bar{\X}_l \right ]_{j_l}= \left [\bar{\Y}_l \right ]_k \cdot \left [ F^{[p]}(\X) \right ]_{k,j_1,\ldots,j_p} \cdot \prod_{l \neq i=1}^{p} \left [\bar{\X}_i \right ]_{j_i}
$$
for given input $\X \in \R^n$, output adjoint $\bar{\Y}_{l} \in \R^m$ and 
input tangents or adjoints $\bar{\X}_{i} \in \R^n$ yielding the input adjoint
$\bar{\X}_l \in \R^n.$
They allow for $F^{[p]}$ to be evaluated with a relative 
computational cost of $\mathcal{O}(m \cdot n^{p-1})$ 
by letting 
$\bar{\Y}_l$ and the $\bar{\X}_i,$ $l \neq i=1,\ldots,p,$ 
range independently over the Cartesian basis vectors in $\R^m$ and $\R^n,$
respectively. Again, potential sparsity of $F^{[p]}$ should be exploited.
The prime use case for adjoints 
is the computation of gradients ($p=1$ and $m=1$) with a relative 
computational cost of $\mathcal{O}(1).$ This method is also known as 
``back-propagation'' in the context of deep neural networks. 

Equation~(\ref{cr3}) yields
$$
\bar{x}_l= \bar{\Y}^T_l \cdot \prod_{i=2}^{q} F_{i}^{[1]} \cdot F_{1}^{[p]} \cdot \prod_{l \neq j=1}^{p} \bar{x}_j 
$$
where $\bar{x}_{j} \in \R$ for $j=1,\ldots,p$ and $\bar{\Y}_l \in \R^{|\tilde{C}|}.$ A solution for $\bar{\Y}_l={\bf 1} \in \R^{|\tilde{C}|}$ and
$\bar{x}_{j}=1,$ $l \neq j=1,\ldots,p,$ implies a solution of 
{\sc Chain Rule Differentiation}. The number of additions performed on top
of the scalar multiplications is invariant and equal to $|\tilde{C}|-1.$
The corresponding
{\sc Adjoint Differentiation} problem becomes at least as hard as 
{\sc Chain Rule Differentiation}. 

Refer to the literature on algorithmic differentiation for a comprehensive
discussion of first- and higher-order tangents and adjoints.

\section{Conclusion}

The understanding of the computational complexity of discrete problems
is a crucial prerequisite for the development of effective algorithms 
for their (approximate) solution. 
The efficient computation of first and higher derivatives of numerical 
simulations has been both major challenge and fundamental motivation of 
research and development
within the intersection of numerical analysis and 
theoretical computer science for many decades. 
Algorithmic progress has largely been based on the assumption about
{\sc Chain Rule Differentiation} being computationally intractable. A formal
proof has been missing so far. This gap in the theoretical foundations of algorithmic differentiation is filled by this paper.

A substantial body of known results on 
discrete problems in algorithmic differentiation exists. It comprises, 
for example, 
coloring methods for the compression of sparse derivative tensors 
\cite{Gebremedhin2005WCI}, algorithms for efficient data flow reversal 
in adjoint simulations \cite{Griewank1992ALG} and
elimination methods on dags \cite{Naumann2004OAo}. 
Refer to the proceedings of so far seven international 
conferences on algorithmic differentiation, e.g, \cite{Bischof2008AiA,Christianson2018SIA,Forth2012RAi}, for a comprehensive 
survey of numerical, e.g, \cite{Griewank2018Pls}, discrete, e.g. \cite{Chen2012AIP}, and
implementation, e.g, \cite{Pascual2018Mla},
issues as well as for reports on a large number
number of successful applications in computational science and engineering, e.g,
\cite{Grabner2008ADf,Hascoet2018SaA,Ozkaya2012AoA}.
Research and development efforts due to the recent increase in interest in
artificial intelligence and machine learning \cite{Goodfellow-et-al-2016}
are expected to benefit tremendously from this rich collection of results.
The algorithmic differentiation community's web portal {\tt www.autodiff.org}
contains further links in addition to a comprehensive bibliography on the 
subject.


\begin{thebibliography}{10}

\bibitem{Baur1983TCo}
W.~Baur and V.~Strassen.
\newblock The complexity of partial derivatives.
\newblock {\em Theoretical Computer Science}, 22:317--330, 1983.

\bibitem{Bellman1957DP}
R.~Bellman.
\newblock {\em {Dynamic Programming}}.
\newblock Dover Publications, 1957.

\bibitem{Bischof2008AiA}
C.~Bischof, M.~B{\"u}cker, P.~Hovland, U.~Naumann, and J.~Utke, editors.
\newblock {\em Advances in Automatic Differentiation}, volume~64 of {\em
  Lecture Notes in Computational Science and Engineering}.
\newblock Springer, Berlin, 2008.

\bibitem{Chen2012AIP}
J.~Chen, P.~Hovland, T.~Munson, and J.~Utke.
\newblock An integer programming approach to optimal derivative accumulation.
\newblock In {\em \cite{Forth2012RAi}}.

\bibitem{Christianson2018SIA}
B.~Christianson, S.~Forth, and A.~Griewank, editors.
\newblock {\em SPECIAL ISSUE: Advances in Algorithmic Differentiation.}, volume
  33:4--6 of {\em Optimization Methods and Software}.
\newblock Taylor \& Francis, 2018.

\bibitem{Forth2012RAi}
S.~Forth, P.~Hovland, E.~Phipps, J.~Utke, and A.~Walther, editors.
\newblock {\em Recent Advances in Algorithmic Differentiation}, volume~87 of
  {\em Lecture Notes in Computational Science and Engineering}.
\newblock Springer, Berlin, 2012.

\bibitem{Garey1979CaI}
M.~Garey and D.~Johnson.
\newblock {\em Computers and Intractability: A Guide to the Theory of
  NP-Completeness (Series of Books in the Mathematical Sciences)}.
\newblock W. H. Freeman, first edition edition, 1979.

\bibitem{Gauss1801DA}
C.~Gauss and tr. A.~Clarke.
\newblock {\em Disquisitiones Arithmeticae}.
\newblock Yale University Press, 1965.

\bibitem{Gebremedhin2005WCI}
A.~Gebremedhin, F.~Manne, and A.~Pothen.
\newblock What color is your {J}acobian? {G}raph coloring for computing
  derivatives.
\newblock {\em SIAM Review}, 47(4):629--705, 2005.

\bibitem{Godbole1973}
S.~{Godbole}.
\newblock On efficient computation of matrix chain products.
\newblock {\em IEEE Transactions on Computers}, C-22(9):864--866, Sep. 1973.

\bibitem{Goodfellow-et-al-2016}
I.~Goodfellow, Y.~Bengio, and A.~Courville.
\newblock {\em Deep Learning}.
\newblock MIT Press, 2016.

\bibitem{Grabner2008ADf}
M.~Grabner, T.~Pock, T.~Gross, and B.~Kainz.
\newblock Automatic differentiation for {GPU}-accelerated {2D/3D} registration.
\newblock In {\em \cite{Bischof2008AiA}}.

\bibitem{Griewank1992ALG}
A.~Griewank.
\newblock Achieving logarithmic growth of temporal and spatial complexity in
  reverse automatic differentiation.
\newblock {\em Optimization Methods and Software}, 1:35--54, 1992.

\bibitem{Griewank2018Pls}
A.~Griewank, T.~Streubel, L.~Lehmann, M.~Radons, and R.~Hasenfelder.
\newblock Piecewise linear secant approximation via algorithmic piecewise
  differentiation.
\newblock In {\em \cite{Christianson2018SIA}}.

\bibitem{Griewank2008EDP}
A.~Griewank and A.~Walther.
\newblock {\em Evaluating Derivatives: {P}rinciples and Techniques of
  Algorithmic Differentiation}.
\newblock Number 105 in Other Titles in Applied Mathematics. SIAM,
  Philadelphia, PA, 2nd edition, 2008.

\bibitem{Hascoet2018SaA}
L.~Hasco{\"e}t and M.~Morlighem.
\newblock Source-to-source adjoint algorithmic differentiation of an ice sheet
  model written in {C}.
\newblock In {\em \cite{Christianson2018SIA}}.

\bibitem{Karp1972RAC}
R.~Karp.
\newblock Reducibility among combinatorial problems.
\newblock In R.~Miller and J.~Thatcher, editors, {\em Complexity of Computer
  Computations}, pages 85--103. 1972.

\bibitem{Naumann2004OAo}
U.~Naumann.
\newblock Optimal accumulation of {J}acobian matrices by elimination methods on
  the dual computational graph.
\newblock {\em Mathematical Programming, Ser.~A}, 99(3):399--421, 2004.

\bibitem{Naumann2008OJa}
U.~Naumann.
\newblock Optimal {J}acobian accumulation is {NP}-complete.
\newblock {\em Mathematical Programming, Ser.~A}, 112(2):427--441, 2008.

\bibitem{Naumann2012TAo}
U.~Naumann.
\newblock {\em The Art of Differentiating Computer Programs: {A}n Introduction
  to Algorithmic Differentiation}.
\newblock Number~24 in Software, Environments, and Tools. SIAM, Philadelphia,
  PA, 2012.

\bibitem{Ozkaya2012AoA}
E.~{\"O}zkaya, A.~Nemili, and N.~Gauger.
\newblock Application of automatic differentiation to an incompressible {URANS}
  solver.
\newblock In {\em \cite{Forth2012RAi}}.

\bibitem{Pascual2018Mla}
V.~Pascual and L.~Hasco{\"e}t.
\newblock Mixed-language automatic differentiation.
\newblock In {\em \cite{Christianson2018SIA}}.

\bibitem{Rall1981ADT}
L.~Rall.
\newblock {\em Automatic Differentiation: {T}echniques and Applications},
  volume 120 of {\em Lecture Notes in Computer Science}.
\newblock Springer, Berlin, 1981.

\bibitem{Wengert1964ASA}
R.~Wengert.
\newblock A simple automatic derivative evaluation program.
\newblock {\em Communications of the {ACM}}, 7(8):463--464, 1964.

\end{thebibliography}
\end{document}